\documentclass[conference]{IEEEtran}

\usepackage{epsfig}
\usepackage{amssymb}
\usepackage{amsmath}
\usepackage{amsfonts}
\usepackage{algorithm}
\usepackage{algorithmic}
\usepackage{multirow}
\usepackage[font=sf, labelfont={sf,bf}, margin=1cm]{caption}
\usepackage{amsthm}
\usepackage{url}

\ifCLASSINFOpdf
\else
\fi
\usepackage{graphicx}
\usepackage{amssymb}
\usepackage{amsmath}
\usepackage{amsfonts}
\usepackage{algorithm}
\usepackage{algorithmic}
\usepackage{subfigure}

\begin{document}
%
\title{A Probabilistic View of Neighborhood-based Recommendation Methods}

\author{\IEEEauthorblockN{Jun Wang}
\IEEEauthorblockA{ University of Luxembourg \\
jun.wang@uni.lu}
\and
\IEEEauthorblockN{Qiang Tang}
\IEEEauthorblockA{ University of Luxembourg\\
tonyrhul@gmail.com}}

\maketitle
\begin{abstract}
Probabilistic graphic model is an elegant framework to compactly present complex real-world observations by modeling uncertainty and logical flow (conditionally independent factors). In this paper, we present a probabilistic framework of neighborhood-based recommendation methods (PNBM) in which \emph{similarity} is regarded as an unobserved factor. Thus, PNBM leads the estimation of user preference to maximizing a  posterior over \emph{similarity}. We further introduce a novel multi-layer \emph{similarity} descriptor which models and learns the joint influence of various features under PNBM, and name the new framework MPNBM. Empirical results on real-world datasets show that MPNBM allows very accurate estimation of user preferences.
\end{abstract}

\section{Introduction}
\label{introduction}
\begin{figure}[t]
\centering
\includegraphics[height=2.0in, width=2.4in]{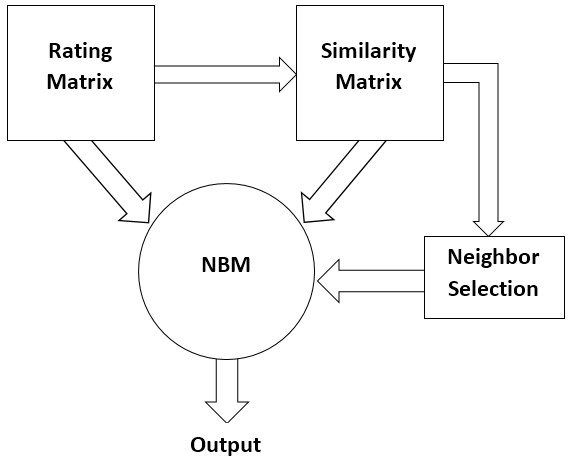}
\caption{ A general structure of NBM}
\label{strnbm}
\end{figure}

Collaborative filtering, which leverages user history information to predict users' unknown preference, is one of the most successful techniques to build recommender systems \cite{su2009survey}. Matrix factorization (MF) \cite{koren2009matrix} and neighborhood-based methods (NBMs) \cite{desrosiers2011comprehensive} are two representative approaches. MF family attracts more attention due to its ability of modeling influence of various features (e.g. \cite{zheng2015incorporating, yang2016learning, koren2008factorization}), thus to improve accuracy. However, it is difficult to provide explainable recommendation results. NBM family, shown as Fig. \ref{strnbm}, is very popular mainly due to the fact that it naturally explains recommendation results (e.g. An item which is similar with what you bought before). \emph{Similarity} serves as the basis of weighting neighbors which is crucial to the accuracy of NBM recommender systems. However, existing \emph{similarity} computation scheme is incapable of capturing influence from different features which hampers further polishing \emph{similarity} to improve accuracy. In this paper, we first present a basic probabilistic framework of NBM family (PNBM) which leads learning \emph{similarity} to a regression problem. Then we introduce a novel multi-layer \emph{similarity} descriptor which models and learns the joint influence of different features under PNBM.

\subsection{Related Work}
\label{relwork1}
Commonly, NBMs are divided into two classes \cite{desrosiers2011comprehensive}. One is user-based approach which predicts the rating that a user will assign to an unrated item by referring to other users who are similar to this user. The other is item-based approach which estimates a user's preference to an unrated item based on other items that are similar to this unrated item. The two approaches follow the same principle.

With respect to NBM, researches have mainly focused on \emph{similarity} computation schemes \cite{desrosiers2011comprehensive} and neighbor selection strategies \cite{adamopoulos2014over}.  \emph{Similarity} also serves as the basis for neighbor selection, thus we concentrate upon \emph{similarity} in this paper. Generally, there are two main approaches to compute \emph{similarity}. One introduces different kinds of correlation coefficients as \emph{similarity} \cite{desrosiers2011comprehensive}, such as Pearson and Cosine correlations. However, some researchers argue that such kind of methods isolate the relations between two items without leveraging global information. The other approach learns \emph{similarity} via regression models. \cite{bell2007scalable, toscher2008improved} introduce a way to learn similarity by minimizing mean squared error between observed ratings and their corresponding estimation. \cite{toscher2008improved} factors similarity matrix via low-rank approximations. \cite{bell2007modeling} presents a weighted  error function which gives more weight to the users who rated items most similar to the estimated item.  \cite{ning2011slim,rendle2009bpr} simplify standard neighborhood-based models to a simple linear regression problem for top-$N$ recommendation based on binary databases.

A number of probabilistic models have been introduced to collaborative filtering. However, only a very small portion of them are NBM related.
 \cite{rendle2009bpr} presents a generic Bayesian personalized ranking framework which is optimized for the area under ROC (AUC) metric. \cite{yu2004probabilistic} introduces a probabilistic memory-based collaborative filtering method in which they use a mixture Gaussian model built on the basis of a set of user profiles and use the posterior distribution of user ratings for prediction.  \cite{defazio2012graphical} builds a Markov network using Pearson-correlation NBM as basis. People also place probabilistic prior assumptions to observations to model uncertainty. Such as, \cite{guo2013novel} places Dirichlet distribution on the absolute value of rating difference. \cite{adamopoulos2014over} uses different probabilistic density functions to sample neighbors from a predefined similarity matrix (vectors).

Unfortunately, these models are incapable of representing complex features, and none of them discusses NBM family itself from a Bayesian perspective.
\subsection{Contribution}

In this paper, we present a probabilistic (Bayesian) framework of NBM family, and our contribution is twofold.

\begin{itemize}
\item First, we present a general graphical model of NBM family (PNBM) which leads the estimation of user preference to maximizing a posterior over \emph{similarity}.
\item Then, we introduce a novel multi-layer \emph{similarity} descriptor which  is capable of modeling and learning the joint influence of various features (e.g. rating, text, genre) under PNBM, and we name the new framework as MPNBM.
\end{itemize}

MPNBM is evaluated on three popular real-world datasets via root-mean-square-error (RMSE) metric. Empirical results show that MPNBM consistently outperform state-of-art approaches  on the datasets we choose.

\section{Preliminary}
\label{preliminary}
Suppose we have a data set organized in form of $User \times Item$ matrix $R \in \mathbb{R}^{N\times M}$, it contains $N$ users and $M$ items. $S \in \mathbb{R}^{M\times M}$ is item \emph{similarity} matrix, $s_{ij}$ denotes similarity between item $i$ and $j$, we further assume $s_{ij}=s_{ji}$. $I \in \mathbb{B}^{N\times M}$ presents indicator matrix, and $ \mathbb{B} =  \{ 0,1\}$. $I_{ui}=1$ if user $u$ rated item $i$, otherwise $I_{ui}=0$. $R^{>0} \subset  R $ denotes all the observed ratings.

So far, many neighborhood-based methods have been proposed, as surveyed in \cite{desrosiers2011comprehensive}. For simplicity, we take a variant of mean-centering NBM \cite{resnick1994grouplens} as instance throughout the paper. The predication formula is defined in Equation (\ref{myeq:mc}).
\begin{equation}\label{myeq:mc}
 \hat{r}_{ui}' = \bar{r}_{i} + \frac{\sum_{j\in \mathcal{I} \backslash \{i\}}s_{ij}(r'_{uj}-\bar{r}_j)I_{uj}}{\sum_{j \in \mathcal{I} \backslash \{i\}}|s_{ij}|I_{uj}}
\end{equation}
where  $r'_{uj}$ is rating score that user $u$ gave to item $j$. $\hat{r}_{ui}'$ denotes the estimation of user $u$'s preference on item $i$. $\bar{r}_i$ is the mean value of all the ratings given to item $i$. $\mathcal{I}$ presents a set containing all the items.


For further simplicity,
Equation (\ref{myeq:mc}) is transformed into a vectorization form:
\begin{equation}\label{myeq:basic}
\hat{r}_{ui} = \frac{\sum_{j\in \mathcal{I} \backslash \{i\}}s_{ij}r_{uj}}{\sum_{j \in \mathcal{I} \backslash \{i\}}|s_{ij}|I_{uj}}=\frac{S_iR_u^{-}}{|S_i|I_u^{-}}
\end{equation}
where $\hat{r}_{ui} = \hat{r}_{ui}'-\bar{r}_{i}$ and $r_{ui}=(r'_{ui}-\bar{r}_i)I_{ui}$.  $S_{i} \in \mathbb{R}^{1\times M}$ denotes \emph{similarity} vector corresponding to item $i$ and  $R_{u} \in \mathbb{R}^{N\times 1}$ represents rating vector of user $u$. The multiplication $S_{i}R_{u}$ denotes the inner product of the two vectors. $I_{u} \in \mathbb{B}^{N\times 1}$ is an indicator vector of user $u$. The symbol $\cdot ^-$ means a vector that does not contain an item which is being predicted. For example, with regard to Equation (\ref{myeq:basic}), $R_{u}^-$ denotes a vector does not contain $r_{ui}$.  Moreover, we assume the testing set is excluded from the training set, when we predict $\hat{r}_{ui}$ in the testing set, $r_{ui}$ in the training set is always zero.

\section{Probabilistic Framework of NBM}
\label{asimple}

\begin{figure}[ht!]
\centering
\includegraphics[height=2.3in, width=2.4in]{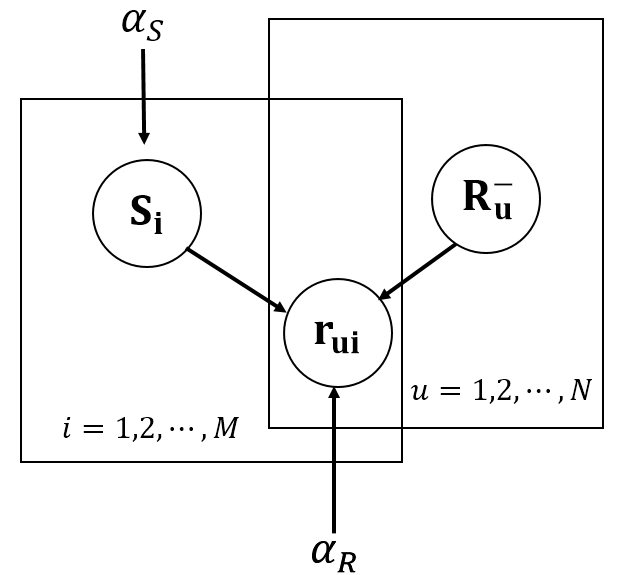}
\caption{Graphical model of PNBM}
\label{gpnbm}
\end{figure}

In this section, we present a probabilistic graphical model of NBMs (PNBM), shown in Fig. \ref{gpnbm}. It is a Bayesian network which describes the following factorization:

\begin{equation}\label{myeq:pnbm}
\begin{split}
p(S_i, R_u^{-}, & r_{ui}, \alpha_S, \alpha_R) = \\ & p(R_u^{-})p(\alpha_S)p(\alpha_R)p(r_{ui}|S_i,R_u^{-},\alpha_R)p(S_i|\alpha_S)
\end{split}
\end{equation}

In our context, placing prior distribution on hyper-parameters $\Theta \{ \alpha_S, \alpha_R \}$ does not significantly improve accuracy while dramatically increasing time complexity. For the sake of simplicity and reduction of time complexity, we simply  let $p(\alpha_S), \ p(\alpha_R)$ be constant, and $p(R_u^{-})$ is also constant. So we can simplify Equation (\ref{myeq:pnbm}) to

\begin{equation}\label{myeq:pnbm2}
p(S_i, R_u^{-},  r_{ui}, \alpha_S, \alpha_R) \propto  p(r_{ui}|S_i,R_u^{-},\alpha_R)p(S_i|\alpha_S)
\end{equation}


We introduce a general Gaussian distribution (but not limited to, other distribution can be also applied to. It depends on real-world context.) to density function $p(*)$ which naturally leads to a sum-of-square-error.

More specifically, assume that an item's similarity vector $S_i$ is independent from those of other items, and $S_i$ is sampled from a mean-zero spherical Gaussian distribution. Thus we have
\begin{equation}\label{myeq:gsim}
p(S|\alpha_{S})=\prod_{i=1}^{M} \mathcal{N}(S_{i}|0, \alpha_{S}^{-1}\mathbf{I})
\end{equation}
where $\mathcal{N}(x|\mu,\alpha^{-1})$ denotes the Gaussian distribution for $x$ with mean $\mu$ and precision $\alpha$.
We also assume that ratings are independent with each other. Combine with Equation (\ref{myeq:basic}), we have following
 \begin{equation}\label{myeq:grate}
p(R^{>0}|S,R^-,\alpha_{R}) = \prod_{i=1}^{M}\prod_{u=1}^{N}[\mathcal{N}(r_{ui}|\frac{S_{i}R_{u}^-}{|S_{i}|I_{u}^-}, \alpha_{R}^{-1}) ]^{I_{ui}}
\end{equation}

\section{Multi-layer \emph{Similarity} Descriptor}
\label{cpnbm}
\begin{figure}[t]
\centering
\includegraphics[height=2.0in, width=2.4in]{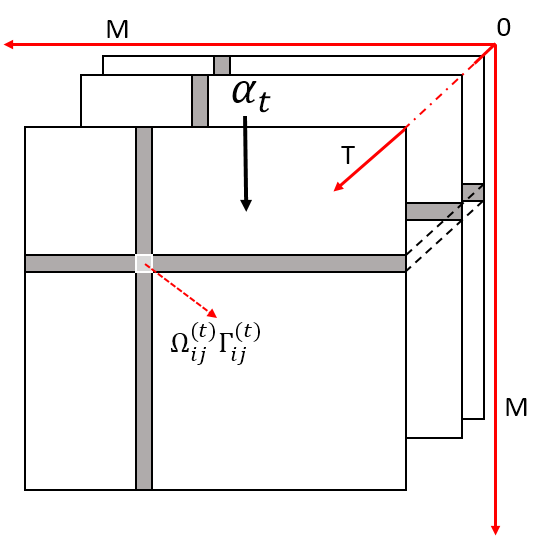}
\caption{Multi-layer \emph{similarity} descriptor. Each layer models an influence generated from features. }
\label{g:mlsd}
\end{figure}

In Section \ref{asimple}, we introduced a general probabilistic (Bayesian) NBM framework which is simple and straightforward. However, like other similarity computation methods, PNBM falls short in  feature representation which extremely limits the accuracy improvement.  In this section, we present a multi-layer \emph{similarity} descriptor (MLSD, shown in Fig.\ref{g:mlsd} ) which is able to model and learn the joint influence of various features (e.g. ratings, text, genre, time). MLSD is mathematically defined as

\begin{equation}\label{myeq:geneconstrainsim}
 \mathcal{S} =\sum_{t=1}^{T}\phi^{(t)}(\Omega^{(t)} \circ \Gamma^{(t)})
\end{equation}
where $\Gamma^{(t)}\in \mathbb{R}^{M\times M}$ denotes the \emph{similarity} basis at $t$-th layer. $\Omega^{(t)} \in \mathbb{R}^{M\times M}$ is $\Gamma^{(t)}$'s constraint matrix which presents an influence of observed features.  For example, it can present the similarity of text description ( or time closeness ) between any two-item. Note that different influences may be generated from the same feature. $\phi^{(t)}$ denotes the importance of the feature-influence modeled at layer $t$. $T$ is the number of layers (influence) employed to model \emph{similarity}. In this paper, we don't require $\sum_{t=1}^{T}\phi^{(t)}=1$, since we always have a normalization factor $|S_i|I_u^-$ in the prediction equation, i.e. Equation (\ref{myeq:basic}). $A \circ B$ denotes point-wise product operation (Hadamard product) on matrices $A$ and $B$, e.g.
$$\begin{pmatrix}
a_{11} &  a_{12}\\
a_{21} &  a_{22}
\end{pmatrix} \circ
\begin{pmatrix}
b_{11} &  b_{12}\\
b_{21} &  b_{22}
\end{pmatrix} =
\begin{pmatrix}
a_{11}b_{11} &  a_{12}b_{12}\\
a_{21}b_{11}&  a_{22}b_{22}
\end{pmatrix} $$

\begin{figure}[t]
\centering
\includegraphics[height=2.2in, width=3.25in]{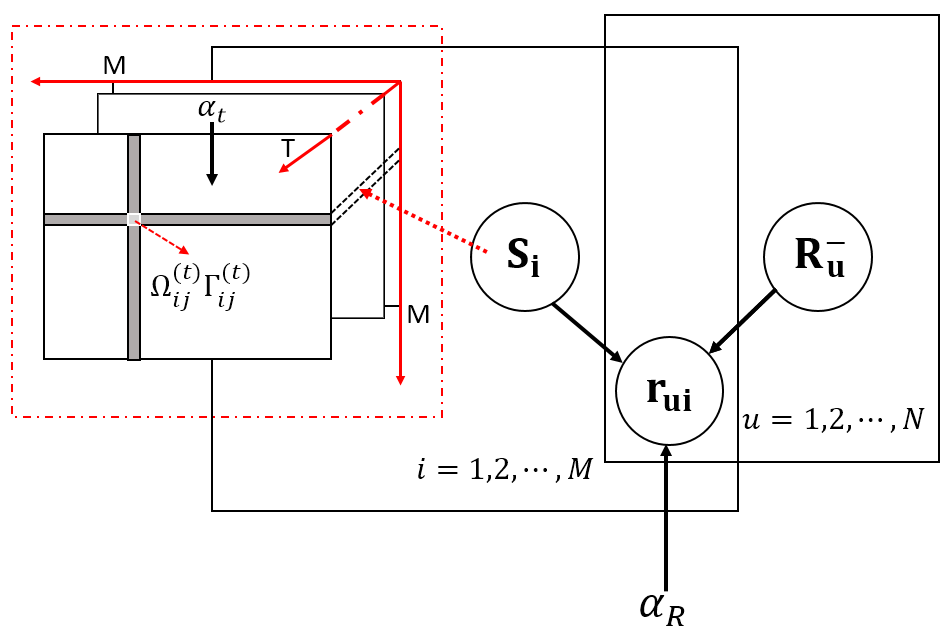}
\caption{Graphical model of MPNBM.  Note that only black solid arrows ($\rightarrow$) denote the dependency flow in the graphical model.}
\label{g:mpnbm}
\end{figure}

MLSD can be smoothly integrated into PNBM, shown in Fig. \ref{g:mpnbm}, named MPNBM. The Bayesian network is mathematically describe as
\begin{equation}\label{myeq:mpnbm}
\begin{split}
p(R_u^{-}, &  r_{ui}, \Gamma^{(t)}_{i}, \Omega^{(t)}_{i}, \alpha_t, \alpha_R) \propto \\
& p(r_{ui}|S_i,R_u^{-},\alpha_R)\prod_{t=1}^{T}  p(\Omega^{(t)}_{i} \circ \Gamma^{(t)}_{i} |\alpha_t)
\end{split}
\end{equation}
where $ \mathcal{S}_i =\sum_{t=1}^{T}\phi^{(t)}(\Omega^{(t)}_i \circ \Gamma^{(t)}_i)$.
Follow the same assumptions in Section \ref{asimple}, we define the prior of layer $t$   ( $\Omega^{(t)}_i \circ \Gamma^{(t)}$) as
\begin{equation}\label{myeq:mulsim}
 p(\Omega^{(t)} \circ \Gamma^{(t)}|\alpha_{t})=\prod_{i=1}^{M} \mathcal{N}(\Omega^{(t)}_{i} \circ \Gamma^{(t)}_{i}|0, \alpha_{t}^{-1}\mathbf{I})
\end{equation}
And we have the conditional distribution over observed ratings defined as
\begin{equation}\label{myeq:rmul}
\begin{split}
p(R^{>0}| & \Omega^{(t)},  \Gamma^{(t)}, R^-, \alpha_R) = \\
& \prod_{i=1}^{M}\prod_{u=1}^{N}\mathcal{N}[(r_{ui}|\frac{(\sum_{t=1}^{T}\phi^{(t)}\Omega^{(t)} \circ \Gamma^{(t)})R_u^{-}}{(\sum_{t=1}^{T}\phi^{(t)}\Omega^{(t)} \circ \Gamma^{(t)})I_u^{-}}, \alpha_R^{-1})]^{I_{ui}}
\end{split}
\end{equation}

\section{Maximum a Posterior}
\label{map-1}
PNBM is a specific case of MPNBM, which only has one layer with constraint-matrix set to 1. In this section, we take MPNBM as example to present how we optimize \emph{similarity} via maximizing a posterior.

The log of Bayesian network defined in Equation (\ref{myeq:mpnbm}) is given by
\begin{equation}\label{myeq:logmpnbm}
\begin{split}
\log p( & R_u^{-}, r_{ui}, \Gamma^{(t)}_{i}, \Omega^{(t)}_{i}, \alpha_t, \alpha_R) \propto \\
& \log p(r_{ui}|S_i,R_u^{-},\alpha_R) + \sum_{t=1}^{T}  p(\Omega^{(t)}_{i} \circ \Gamma^{(t)}_{i} |\alpha_t)
\end{split}
\end{equation}

In fact, it defines the posterior distribution over \emph{similarity}.
Combine it with Equation (\ref{myeq:mulsim}) and Equation (\ref{myeq:rmul}), we have

\begin{equation}\label{myeq:logmpnbm2}
\begin{split}
-  \log &  p(R^{-}, R, \Gamma^{(t)}_{i}, \Omega^{(t)}_{i}, \alpha_t, \alpha_R) \propto \\
& \frac{\alpha_R}{2} \sum_{i=1}^M\sum_{u=1}^N(r_{ui}-\frac{S_iR_u^-}{|S_i|I_u^-})^2+\frac{\alpha_t}{2}\sum_{t=1}^{T}\sum_{i=1}^M(||\Omega^{(t)}_{i} \circ \Gamma^{(t)}_{i}||_2)\\
+& M^2 \sum_{i=1}^T\log \frac{\alpha_t}{\sqrt{2\pi}} + \log \frac{\alpha_R}{\sqrt{2\pi}} \sum_{i=1}^M\sum_{u=1}^NI_{ui}
\end{split}
\end{equation}

Maximizing the above Bayesian network distribution with hyper-parameters being kept fixed is equivalent to minimizing an error function defined as

\begin{equation}\label{myeq:errfunc}
\mathcal{E} = \frac{1}{2} \sum_{i=1}^M\sum_{u=1}^N(r_{ui}-\frac{S_iR_u^-}{|S_i|I_u^-})^2+\sum_{t=1}^{T}\sum_{i=1}^M(\lambda_t ||\Omega^{(t)}_{i} \circ \Gamma^{(t)}_{i}||_2)
\end{equation}
where $\lambda_t = \frac{\alpha_t}{2\alpha_R}$ is the regularization parameter for layer $t$.

A simple linear Gaussian model sometimes makes prediction value fall out of the range of valid rating values. In order to force the predication values to fall into valid range, we pass the linear-Gaussian model through hyperbolic tangent function $h(x)= \frac{e^x-e^{-x}}{e^x+e^{-x}}$
which makes prediction values be in range of [-1,1].
We map the centralized ratings to range [-1, 1] with Equation (\ref{myeq:mapf}).
\begin{equation}\label{myeq:mapf}
t(x)= \frac{x-\frac{max_x+min_x}{2}}{max_x-\frac{max_x+min_x}{2}}
\end{equation}
where $max_x$ and $min_x$ are the max and min value of ratings, respectively. Since the ratings are centralized by their corresponding mean value, we always have $max_x>0$ and $min_x<0$. As a result, the range of valid rating value align with the estimation produced by our models.

The conditional distribution of observed ratings  becomes
\begin{equation}\label{myeq:gratetanh}
p(R^{>0}|S,R^-,\alpha_{R}) = \prod_{i=1}^{M}\prod_{u=1}^{N}[\mathcal{N}(r_{ui}|h(\frac{S_{i}R_{u}^-}{|S_{i}|I_{u}^-}), \alpha_{R}^{-1}) ]^{I_{ui}}
\end{equation}

We adopt stochastic gradient descent (SGD) as learning algorithm to train latent factors, shown in Algorithm \ref{alg:sgd}.

\begin{algorithm}[h]
   \caption{Training via Stochastic Gradient Descent}
   \label{alg:sgd}
\begin{algorithmic}
   \STATE {\bfseries Preliminary:} rating matrix $R$, error function.
   \STATE {\bfseries Initialization:} similarity basis $\Gamma^{(t)}$, influence constraint-matrix $\Omega^{(t)}$, influence importance factor $\phi^{(t)}$, learning rate $\beta$, regular parameter $\lambda_{t}$. Note that $ \mathcal{S}_i =\sum_{t=1}^{T}\phi^{(t)}(\Omega^{(t)}_i \circ \Gamma^{(t)}_i)$.
   \STATE $\bullet \ \ $ Training:
   \FOR{$k=1$ {\bfseries to} $K$}
   \STATE $\bullet \ \ $ For each layer ($t$), point-wisely update the similarity basis $\Gamma^{(t)}_{ij}$:
   \STATE $\quad \quad [\Gamma_{ij}^{(t)}]^{new} =[\Gamma_{ij}^{(t)}]^{old}-\beta  e_{ui} \frac{\partial \hat{r}_{ui} }{\partial \Gamma^{(t)} _{ij}}-\beta \lambda_{t}(\Omega^{(t)}_{ij} \circ \Gamma^{(t)}_{ij})$
   \STATE $\bullet \ \ $ where $e_{ui}=\hat{r}_{ui}-r_{ui}$.
   \ENDFOR
   \STATE {\bfseries Prediction:} prediction using Equation (\ref{myeq:mc}) with top-$200$ the most similar neighbors.
\end{algorithmic}
\end{algorithm}

\section{Experiments}

\label{experiments}
\subsection{Description of Data Sets}
In the experiments, we evaluate our models and state-of-art methods over three different data sets, summarized in Table \ref{dataset-table}.

\begin{table}[h!]
\centering
\caption{Data Sets}
\hspace*{-0.3cm}
\begin{tabular}{|c|c|c|c|c|l|} \hline
data set & user\# & item\# & ratings\# &scales & density \\ \hline
ML-20M  & 138,493 & 26,744 &20,000,263&[0.5,5] & 0.54\%\\ \hline
ML-10M  & 69,878 & 1,0677 &10,000,054&[0.5,5] & 1.34\%\\ \hline
Netflix & 32,682 & 13,139 &3,967,477&[1,5] & 0.92\% \\
\hline
Yahoo-R4 & 7,637& 3,791&207,854&[1,5] & 0.72\%\\
\hline\end{tabular}

\label{dataset-table}
\end{table}

ML-20M, ML-10M are data sets provided by MovieLens \cite{ml1020m}. Netflix is a subset sampled from Netflix Prize data set \cite{netflix1} such that each user rated 50-1500 movies, and each movie is rated by 5-1800 users. Yahoo-R4 is a subset of the movie-rating data set provided by the Yahoo Labs Webscope Team \cite{yahoor4} such that each movie are at least rated by 5 users.
\begin{itemize}
\item We use ML-10M, Netflix and Yahoo-R4 to compare the models' accuracy.
\item We also compare each model's accuracy on data sets with different densities. In order to avoid the inherent differences of data sets from different originations, we extract 10 subsets from one single data set (ML-20M) based on the users' rating number. Precisely, each subset has similar amount of users and items, the number of users and items are in range [10000,  15000] and [8000, 20000] respectively. The density of each data set is from 0.28\% to 2.67\%.
\end{itemize}

\begin{figure*}[!ht]
\hspace*{-0.5cm}
\includegraphics[height=2.8in, width=7in]{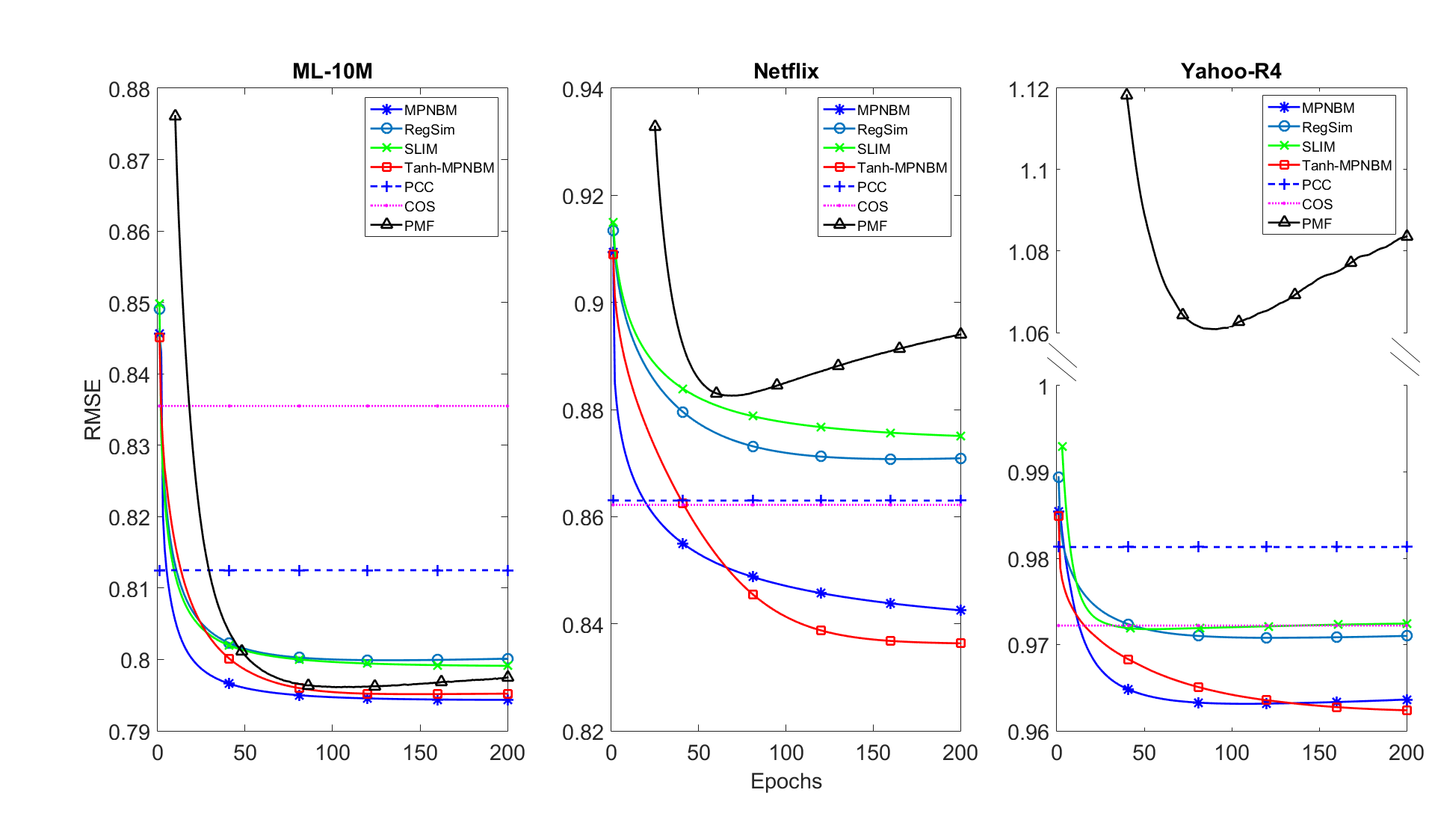}
\caption{ RMSE evaluation on ML-10M, Netflix, Yahoo-R4. The Y-axis displays RMSE value and the X-axis shows the number of epochs (iterations) in the training.}
\label{fig:com}
\end{figure*}

\begin{figure*}[!ht]
\hspace*{-0.2cm}
\includegraphics[height=3in, width=6.8in]{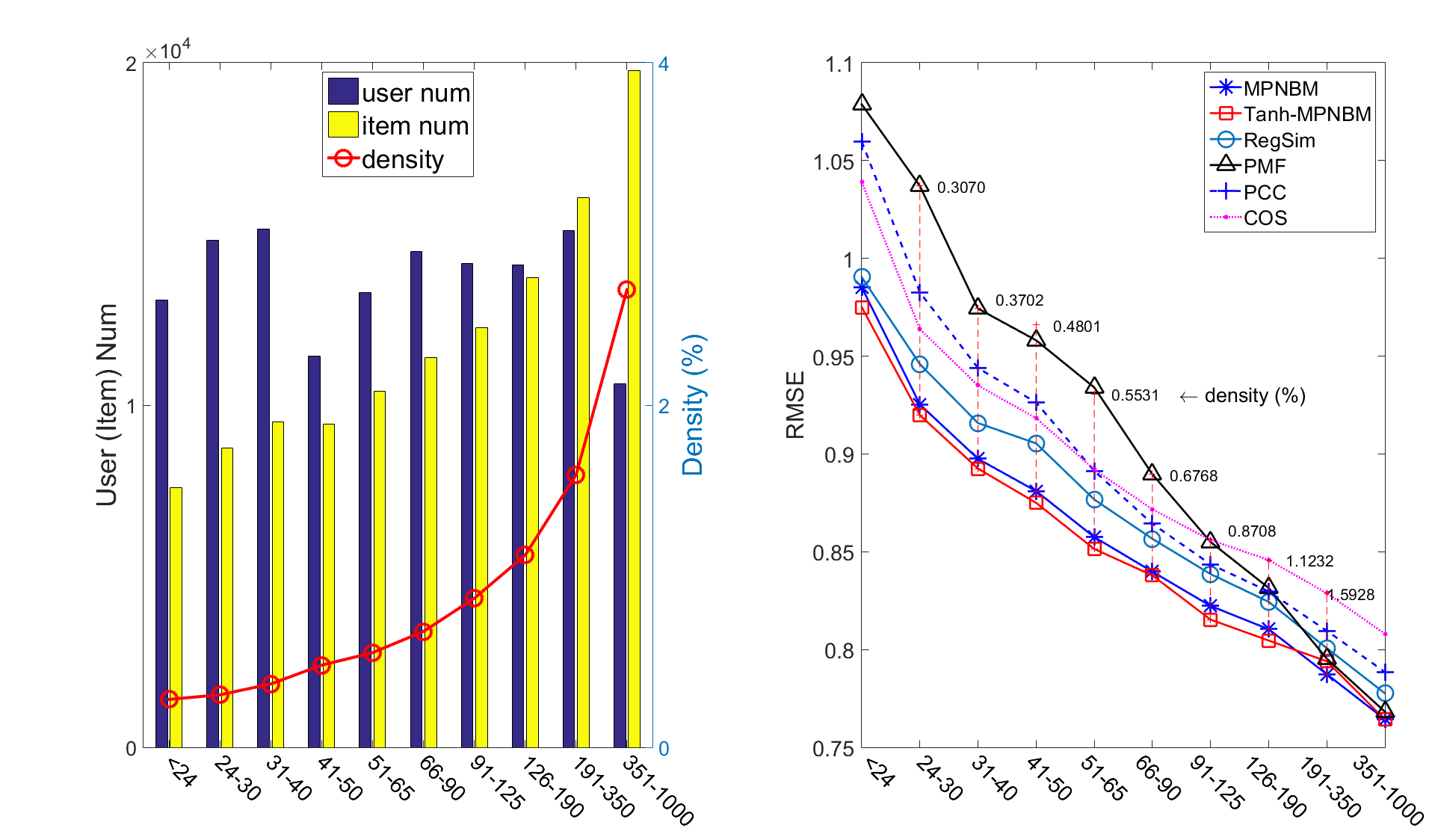}
\caption{ RMSE evaluation with different density. Left panel: Basic information of the 10 subsets extracted from ML-20M. Right panel: The RMSE evaluation on each subset, the Y-axis displays RMSE value. The X-axis of both panels displays the number of rated items per user in each subset.}
\label{diffdensity}
\end{figure*}

\subsection{Models for Comparison}
\label{algfc}
In this paper, the following models are compared:
\begin{itemize}
\item {\bfseries RegSim: } Regression on \emph{similarity}  \cite{toscher2008improved}, a representative work which learns \emph{similarity} via a regression method.
\item {\bfseries SLIM: } Sparse linear methods \cite{ning2011slim}, a regression model for top-$N$ recommendation on binary data set. We extend it to a arbitrary real-value prediction model by placing a jointly Gaussian-Laplace prior on similarity vectors. It has a very similar error function as SLIM,
\begin{equation}
\begin{split}
\mathcal{E}  =   &\frac{1}{2}\sum_{u=1}^{N}\sum_{i=1}^{M} ( r_{ui}-\frac{S_iR_u^-}{|S_i|I_u^-})^{2}I_{ui}  \\
& +\frac{\lambda_S}{2}\sum_{i=1}^{M}||S_{i}||_2-\lambda_S \mu\sum_{i=1}^{M}||S_i||_{1}
 \end{split}
\end{equation}
where $\mu$ is a non-zero mean value of the Gaussian prior.
\item {\bfseries PCC: } NBM using Pearson correlation as \emph{similarity} \cite{resnick1994grouplens}.
\item {\bfseries COS: } NBM using Cosine correlation as \emph{similarity} \cite{desrosiers2011comprehensive}.
\item {\bfseries PMF: } Probabilistic matrix factorization \cite{ruslanpmf}.
\item {\bfseries MPNBM: } In this paper, we exploit influence from ratings as an instance to demonstrate MLSD's ability of modeling various features, thus to improve accuracy. We use a 3-layer \emph{similarity} descriptor in which layer-1 treats latent influence equally with constraint-matrix set to 1; Layer-2 adopts Pearson correlation as constraint-matrix that stresses the influence from those items which either have significant positive correlation or strong negative correlation
with the item under predication; Layer-3 employs Jaccard index to form a constraint-matrix that amplifies the influence from those items which have similar rating history, alleviates the divergence from infrequent-rated items and frequent-rated items. {\bfseries Time Complexity.} The computational time is mainly taken by updating \emph{similarity}. At a single epoch,
approximately $T \cdot \mathcal{L}\cdot \#_u$ similarities are updated, where $\mathcal{L}$ is the size of training set, $\#_u$ is the average rating number per users and $T$ is the number of influence layers. Intuitively, a single epoch takes about 4, 260, 340 seconds on Yahoo-R4, Netflix, ML-10M respectively.
\item {\bfseries Tanh-MPNBM: } The model which we pass MPNBM through hyperbolic tangent function ( detailed in Section \ref{map-1} ).
\end{itemize}

\emph{Experiment setting.} All models are implemented with Matlab, and run on a single core of a Intel (R) Xeon(R) 3.50 GHz machine with 16 GB memory. 

\subsection{Parameters Setting}
For RegSim, MPNBM, Tanh-MPNBM and SLIM,
we empirically choose parameters for each model after a grid search in which $\beta \in \{0.05, 0.1, 0.2, 0.3, 0.4, 0.5\}, \lambda_1=\lambda_2=\lambda_3 \in \{0.01, 0.02, 0.03, 0.04, 0.05, 0.06, 0.1\}$. The finally chosen parameters are summarized in Table \ref{setpara} ($\perp$ indicates a model does not have such a parameter).
\begin{table}[!ht]
\centering
\caption{Parameters Setting for RegSim, MPNBM, Tanh-MPNBM, SLIM.}
\begin{tabular}{|l|c|c|c|}
\hline
           & $\beta$ & $(\lambda_1,\lambda_2,\lambda_3)$ & $(\phi^{(1)},\phi^{(2)},\phi^{(3)})$ \\ \hline
RegSim       & 0.1     & (0.01,$\perp$,$\perp$)           & ($\perp$,$\perp$,$\perp$)                              \\ \hline
MPNBM      & 0.2     & (0.05,0.05,0.05)                 & (3,1,1)                                                \\ \hline
Tanh-MPNBM & 0.4     & (0.05,0.05,0.05)                 & (3,1,1)                                                \\ \hline
SLIM    & 0.4     & (0.02,$\perp$,$\perp$)           & ($\perp$,$\perp$,$\perp$)                              \\ \hline
\end{tabular}

\label{setpara}
\end{table}

For PMF,  we choose latent feature dimension $D=10$ and the momentum of mini-batch SGD $\eta=0.8$. Regularized parameters ($\lambda_P, \lambda_Q$ for user latent factors and latent item factors respectively) and learning rate $\beta$ are set to

\begin{itemize}
\item For Yahoo-R4, $\lambda_P=\lambda_Q=0.05$ and $\beta=0.0005$.
\item For Netflix, $\lambda_P=\lambda_Q=0.002$ and $\beta=0.0002$.
\item For ML-10M, $\lambda_P=\lambda_Q=0.02$ and $\beta=0.0002$.
\item For the first two ML-20M subsets shown in the left panel of Fig. \ref{diffdensity}, $\lambda_P=\lambda_Q=0.01$ and $\beta=0.0002$.
\item For the other eight ML-20M subsets shown in the left panel of Fig. \ref{diffdensity}, $\lambda_P=\lambda_Q=0.02$ and $\beta=0.0002$.
\end{itemize}

For PCC and COS, we use top-200 the most similar neighbors for prediction.

\subsection{Comparison Results}
During the test, we randomly divide each data set into training set (85\%), validation set (5\%) and testing set (10\%). We adopt RMSE for evaluation.
We repeat the experiments 5 times.
\begin{table}[!ht]
\small
\centering
\caption{Accuracy Comparison (The smaller RMSE, the better accuracy for recommendation.). MPNB,  TMPN, denote MPNBM and Tanh-MPNBM respectively. }
\begin{tabular}{|l|c|c|c|c|c|c|}
\hline
\multirow{2}{*}{} & \multicolumn{2}{l|}{Yahoo-R4} & \multicolumn{2}{c|}{Netflix}                & \multicolumn{2}{l|}{ML-10M} \\ \cline{2-7}
                  & RMSE       & INC\%            & RMSE                        & INC\%         & RMSE      & INC\%           \\ \hline
RegSim              & 0.9723     & 0             & 0.8713                      & 0        & 0.8034    & 0           \\ \hline
MPNB             & 0.9641     & \textbf{0.84}    & 0.8425                      & \textbf{3.31} & 0.7941    & \textbf{1.16}   \\ \hline
TMPN      & 0.9629     & \textbf{0.97}    & 0.8363                      & \textbf{4.02} & 0.7955    & \textbf{0.98}   \\ \hline
SLIM           & 0.9725     & -0.02             & 0.8731                      & -0.21         & 0.8004    & 0.37            \\ \hline
PMF               & 1.0608     & -9.1            & 0.8826                      & -1.3         & 0.7957    & 0.96             \\ \hline
PCC               & 0.9813     & -0.93                & \multicolumn{1}{l|}{0.8620} & 1.07             & 0.8121    & -1.08               \\ \hline
COS               & 0.9722     & 0.01             & \multicolumn{1}{l|}{0.8605} &1.24          & 0.8362    & -4.08           \\ \hline
\end{tabular}
\label{accuracy-table-rmse}
\end{table}

\subsubsection{Accuracy}
\label{cpt}
The comparison is performed over :
\begin{itemize}
\item Accuracy on different data sets.
\item Accuracy on different density.
\end{itemize}

Fig. \ref{fig:com}  presents the detail of training on different data sets. Table \ref{accuracy-table-rmse} records the final accuracy  comparison (the training process is conducted by validation set). RegSim is selected as baseline model,  the accuracy improvement of each model is displayed in the INC \% column.

Fig. \ref{diffdensity} shows the accuracy comparison on data sets with different density.  MPNBM and Tanh-MPNBM consistently outperform outperform state-of-art models, especially on those extremely sparse data sets ( which have serious \emph{cold start} problem). For simplicity, we don't draw SLIM on the graph, since SLIM has similar accuracy with RegSim.

\begin{figure}[!ht]
\hspace*{-0.4cm}
\includegraphics[height=2.8in, width=3.7in]{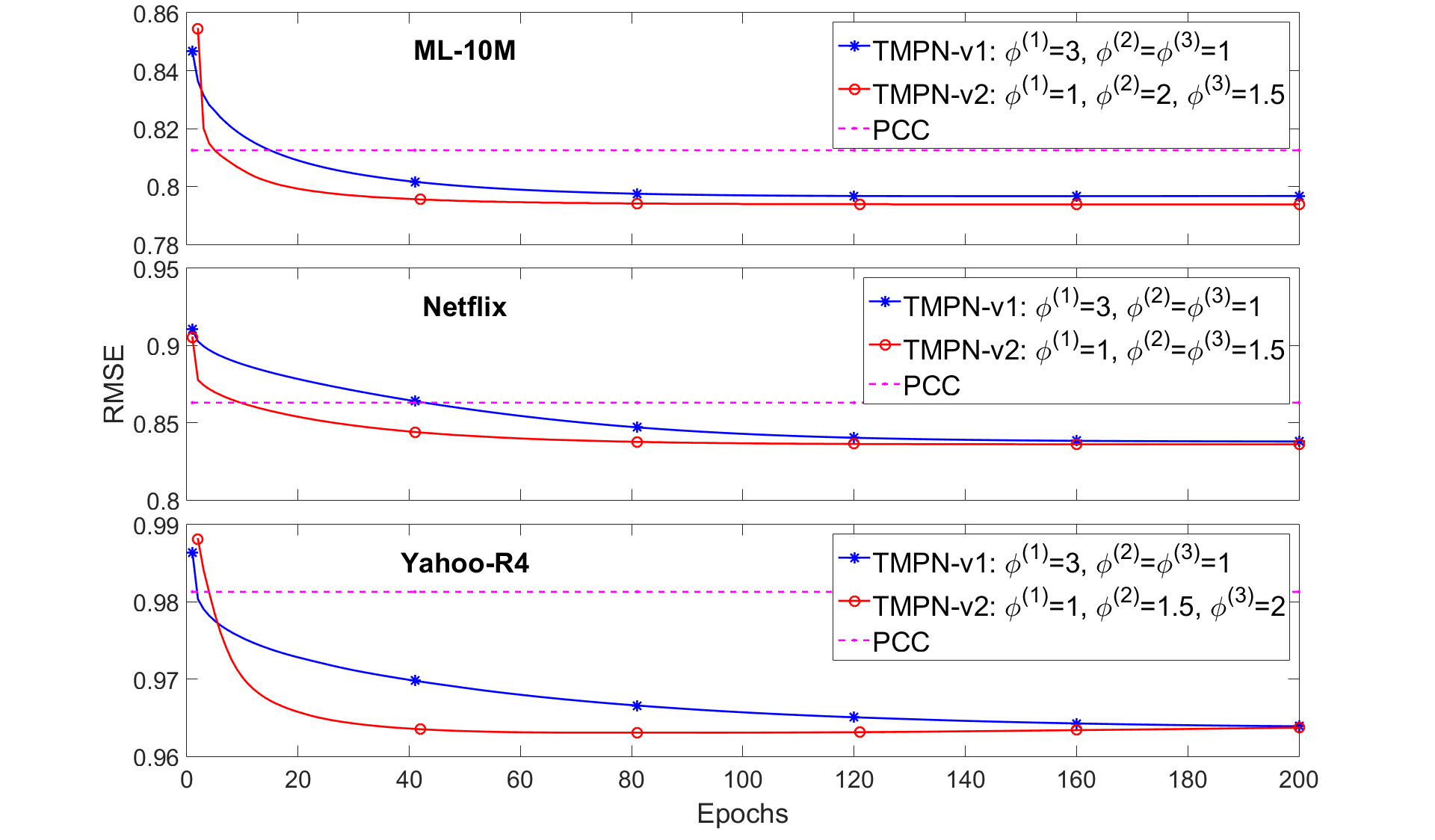}
\caption{ Tanh-MPNBM: Comparison between two strategies of setting $\phi$. }
\label{fig:toy}
\end{figure}

\vspace{0.2cm}

We are also interested in that how the layer importance-factor $\phi$ affects the MPNBM (Tanh-MPNBM). We use two strategies to select parameters $\phi$ for each layer, 1) we consistently choose $\phi^{(1)}=3,\ \phi^{(2)}=\phi^{(3)}=1$ for all the three data sets, named TMPN-V1; 2) letting $\phi^{(t)} \in \{1, 1.5, 2 \}$, we assign higher value to the $\phi$ which corresponding $\Omega$ has lower RMSE, named TMPN-V2.  The comparison is shown in Fig. \ref{fig:toy}, and 1) the accuracy is not significantly influenced, MPNBM (Tanh-MPNBM) is able to balance the influence automatically. 2) Assigning proper weight to $\phi$ according to the RMSE of $\Omega$ results in faster convergence.

\subsubsection{Stability}

Model based approach may easily over fit when increasing the number of parameters under training. The system can be beneficent from the stability of algorithms which is defined by
\begin{itemize}
\item converge speed: the first epoch where a model converges to the local best solution, denoted as $\epsilon$.
\item  ability of models to maintain their best status: the number of epochs that a model stays in the local best solution, denoted as $\zeta$.
\end{itemize}

Table \ref{stable-table} shows the values of $\epsilon$ and $\zeta$ of each model over different data sets.
\begin{table}[h!]
\centering
\caption{Stability Comparison}
\begin{tabular}{|l|c|c|c|c|c|c|}
\hline
\multirow{2}{*}{} & \multicolumn{2}{l|}{Yahoo-R4} & \multicolumn{2}{c|}{Netflix} & \multicolumn{2}{l|}{ML-10M} \\ \cline{2-7}
                  &     $\epsilon$          &    $\zeta $           &              $\epsilon$  &      $\zeta $         &     $\epsilon$           &            $\zeta $  \\ \hline
RegSim              & 86            & 102           & 134           & $\geq$67           & 96            & 91          \\ \hline
MPNBM             & 84            & 59            & \multicolumn{2}{c|}{*}       & 141           & $\geq$60          \\ \hline
Tanh-MPNBM        & 158           & $\geq$43            & 166           & $\geq$35           & 115           & $\geq$86          \\ \hline
SLIM           & 39             & 51             & 186             & $\geq$15            & 150             & $\geq$51           \\ \hline
PMF               & 87            & 6             & 64            & 12           & 93            & 34          \\ \hline
\end{tabular}
\label{stable-table}
\end{table}
In the comparison of stability, we treat RMSE values $x_1 = x_2$, if $|x_1-x_2|\leq 0.0001$. Note that with regard to a model which does not over fit after 200 epochs (value of $\zeta$ prefixed with $\geq$), if the lowest RMSE value appears at least 10 epochs, it is seen as the local best solution. * in Table \ref{stable-table} means a model does not converge after 200 epochs on a data set. e.g, MPNBM does not converge on Netflix data set, also shown in Fig. \ref{fig:com}. The experimental results show that MPNBM and Tanh-MPNBM stay in the local best solution for many ( $>$ 40) epochs which is better than PMF. With regard to converge speed, as shown in Table \ref{stable-table}, it seems that sometimes MPNBM and Tanh-MPNBM do not converge as fast as PMF. In fact, they achieve a considerable accuracy at a much earlier epoch.

\section{Conclusion}
\label{conclusion}
In this paper, we have presented a probabilistic framework of NBM family, and introduced a multi-layer \emph{similarity} descriptor under PNBM which is capable of modeling and learning the joint influence of various features. Our experiments show that MPNBM and Tanh-MPNBM allow accurate and stable estimation of user preferences.

Privacy is a serious problem to recommender systems. Nowadays, applying differential privacy to recommendation algorithms attracts great attention. A common approach is adding noise to data set. Recently, people find that sampling from a posterior distribution achieves some extent of differential privacy ``for free" \cite{wang2015privacy}, and this idea has been already successfully applied to probabilistic matrix factorization \cite{liu2015fast}. Following the same idea, our models can also provide such kind of ``free privacy". We leave a detailed investigation as future work.

\section{Acknowledgments}
Both authors are supported by a CORE (junior track) grant from the National Research Fund, Luxembourg. Qiang Tang is also partially supported by an internal project from University of Luxembourg.

%
\bibliographystyle{abbrv}
\bibliography{sigproc}  

\end{document}